\documentstyle[preprint,aps]{revtex}
\draft
\newcommand{\uni}{~}
 
\begin{document}
 
\title{Effects of epitaxial strain and 
ordering direction on the electronic
properties of (GaSb)$_1$/(InSb)$_1$ and (InAs)$_1$/(InSb)$_1$ superlattices}
\author{S.Picozzi and A. Continenza}
\address{Dipartimento di Fisica \\
Universit\`a degli Studi di L'Aquila, 67010 Coppito (L'Aquila), Italy}
\author{and}
\author{A.J.Freeman}
\address{Department of Physics and Astronomy and Materials Research Center\\
Northwestern University, Evanston, IL 60208-3112 (U.S.A.)\\}
\maketitle
 
\narrowtext
 
\begin{abstract}
 
The structural and electronic properties in
common\--anion \\
(GaSb)$_1$/(InSb)$_1$ and common\--cation (InAs)$_1$/(InSb)$_1$
[111] ordered super\-lattices have been determined using the local density
total energy full potential linearized augmented plane wave method.
The influence of the ordering direction, strain
conditions and atomic substitution on the electronic properties of
technological and experimental interest (such as energy band-gaps
and charge carrier localization in the different sublattices) were
determined. The
results show an appreciable energy band-gap narrowing compared to the
band-gap averaged over the constituent binaries, either
in [001] ordered structures or (more markedly)  in the [111]
systems;  moreover energy band-gaps  show an increasing  trend as
the substrate lattice parameter is decreased.
Finally, the systems examined offer interesting opportunities for
band-gap tuning as a function of the
growth condition (about 0.7~eV in (GaSb)$_1$/(InSb)$_1$ and 0.3~eV in
(InAs)$_1$/(InSb)$_1$).

\end{abstract}

\section{Introduction}
Ternary systems based on
III-V semiconductors (such as disordered alloys  \cite{dj,laks,lemos},
heterostructures \cite{aleoff,christensen,anderson,tit,rubio,az94}
or quantum well systems \cite{vlaev,prins,qw3}) have been the
subject of wide scientific interest and
of accurate theoretical studies, 
since they could be used as fundamental components in a large class of
important devices  (laser diodes
or infrared detectors, to name just a few) \cite{materials}. In the
present work,  we 
focus our attention on superlattices (SL),
whose structural, electronic and transport properties can be opportunely
tuned 
by varying the constituent materials, the strain, the
ordering direction or the layers
 thickness. 
To this end, we have
 examined the properties of interest in
ultrathin [111] ordered SL, specifically in   common-anion
(GaSb)$_1$/(InSb)$_1$ systems
and in  common-cation (InAs)$_1$/(InSb)$_1$ systems, using the
self-consistent all-electron FLAPW method with\-in the density functional
formalism. The systems considered here are now under experimental
investigation (results obtained for strained-layer InSb/GaSb quantum well
\cite{qwapl}, In$_x$Ga$_{1-x}$Sb/GaSb heterostructures \cite{qwjvc}
and InAs$_{1-x}$Sb$_{x}$ alloys \cite{inassb1,inassb2} 
have already been published); at the present time, however,
we are not yet able to compare our predicted results
with experimental values regarding the  SL.
 
The appreciable mismatch  between the lattice parameters of the
binary constituents (5.7~$\%$ in (GaSb)$_1$/(InSb)$_1$ and 6.4~$\%$ in
(InAs)$_1$/(InSb)$_1$)
gives the opportunity to study the effects of the strain conditions
on the SL electronic properties. In analogy with
the common experimental approach \cite{qwapl,ohler,bensaada}
we have considered various growth conditions for the SL, leading
to different strain modes in the structure:
(i) pseudomorphic growth
on a substrate usually constituted by one of the binary constituents
(in which the  lattice constant parallel to the growth plane is
taken equal to that of the bulk
semiconductor composing the substrate) and
(ii) a ``free standing mode'' (FSM), in which
no constraints are imposed on the bond-lengths, leading to relaxed
lattice constants for the binary constituents, both different
from their bulk values.
The  structural parameters for all of the structures considered
have been chosen through total energy minimization
or according to the macroscopic theory of elasticity (MTE) \cite{VDW}.
 
The choice made towards the [111] direction is encouraged by recent
experimental observations \cite{osscupt} that suggested the spontaneous
ordering along this particular axis
(the so called CuPt structure)
shown by some III-V alloys during
vapour-phase growth.
The strong influence of the ordering direction  on the SL electronic
properties is immediately clear if we consider the
ternary SL Brillouin Zones (BZ) as obtained
 from the binary zincblende BZ through folding operations,
which are obviously different for [111] or [001] ordered systems; the
 immediate consequence of this is a noticeable
difference in the electronic properties.
 
In order to study  the dependence of the electronic properties
on the ordering direction, we have studied  the common-anion and
the common-cation systems in the three different growth conditions
both in the CuAu-like
(having [001] direction as growth axis) and CuPt-like structure.
 
Following the model proposed by Wei and Zunger \cite{az89}, we  consider
the SL as obtained first from
an ideal virtual crystal (a common-anion (common-cation) system
having the cation (anion) with  intermediate
properties between the two cations relative to the binary constituents).
We then introduce a perturbative
potential, having a structural part (due to atomic displacements and
relaxation due to epitaxial strain)
and a chemical one (due to the electronegativity difference between the
constituent atoms).
 
In order to separate the  effects due to the two different terms
in the expression of the potential, 
we have studied strained (AC)$_1$/(AC)$_1$-type systems ideally
obtained by monolayer deposition
of the same binary constituent (AC) along
the [111] direction, in which the two different AC
bond-lengths are equal to those in the
equivalent (AC)$_1$/(BC)$_1$ SL.
 
We will discuss the results obtained in this work as follows:
first of all, we will briefly expose  the computational
details and parameters used in the
calculations (Section II); in Section III our structural results
will be reported for all the different systems considered.
In Section IV
we will discuss the electronic properties of the SL, with particular
attention to the quantities of technological interest (such as
energy band-gaps and crystal field splittings) and their dependence on
growth direction, atomic substitution and strain conditions.
We will also discuss the charge density distribution and 
in particular the localization of the charge carriers in the different
constituents sublattices.
Section V summarizes our main results and draws some conclusions.

\section{Method of calculation}

We have determined the properties of the structures considered using
the density functional formalism, within the local density
approximation (LDA) \cite{HK64}
exchange and correlation potential
as parametrized by Hedin\--Lund\-qvist \cite{HL}.
The calculations were performed using
the {\em ab initio} all-electron 
full\--pot\-ential linearized
augmented plane wave method (FLAPW) \cite{FLAPW}. Core electrons
as well as valence ones are treated using a self-consistent
procedure; the shallow Ga $3d$ and In $4d$ states are
considered as valence
states, for which scalar-relativistic effects are included
in the self-consistent calculation, whereas
spin-orbit effects are treated in a perturbative  approach.
For [111] ordered systems, angular momenta up to $l_{max}$ = 6
in the muffin tin spheres (with radius
$R_{\alpha}$ = 2.4 a.u. for all the constituents atoms) and
plane waves with wave vector up to $k_{max}$ = 3.3 a.u. are used, leading to
about 600  basis functions.
 
To perform integrations in reciprocal space,
a set of four special {\bf k} points is chosen in  the
trigonal Brillouin zone (BZ), following the Monk\-horst\--Pack scheme \cite{MP}.
Similar values for these computational parameters
have been used for [001] ordered systems, with the only exceptions
represented by $l_{max}$ = 8 
and a set of
three special {\bf k} points used for the integration over the tetragonal
Brillouin zone. Finally, the Broyden \cite{broy} method is used to
accelerate the convergence in
the self-consistent iterations.
 
\section{Structural properties}
 
The atomic ordering along the [111] direction of a SL grown on a (111) 
substrate gives origin to a trigonal
Bravais lattice with $C^5_{3v}$ (Sch\"oenflies notation) space group
\cite{cryst}.
The unit cell in real space contains 4 atoms and  the origin is taken on a
cation site \cite{magri}.

The free structural
parameters are determined following the
macroscopic theory of elasticity (MTE) \cite{VDW},
taking into account the elastic
properties of the constituent materials, and then compared with those
obtained through total energy minimization.
We observe that in each cell, there are two atoms belonging to the same
chemical species (the two Sb anions in the (GaSb)$_1$/(InSb)$_1$
systems and the two In cations  in the (InAs)$_1$/(InSb)$_1$
systems), which are not equivalent from the coordination point of view.
As an example, we consider the particular case of (GaSb)$_1$/(InSb)$_1$ SL,
having the first Sb$_{Ga}$ bound with three Ga and one In, and the second
Sb$_{In}$ showing a complementary situation.
The total energy minimization procedure that considers all the 
free parameters in the unit cell
as different degrees of freedom is a very onerous computational problem;
this encouraged some simplifications, such as
considering equal bond lengths between
equal atomic species ({\em i.e.,} in the common-anion system, we have chosen
$d_{InSb_{Ga}}\:=\:d_{InSb_{In}}$), thus reducing to three the
number of degrees of freedom
(in-plane lattice parameter and two different bond-lengths). 
Although this simplification is frequently used
in total energy minimization \cite{magri}, one should be
aware that this approximation results in considering an
average over the two different local environments and therefore that
the real elastic structures may slightly differ from our
optimized SLs.

According to MTE, using the same notation as in Ref.\cite{VDW},
the structural parameters for the epilayer
are determined as follows:
\begin{eqnarray}
a^{epi}_{\parallel} &=& a^{subs} \nonumber \\ \nonumber
a^{epi}_{\perp} & = & a^{epi}\:\left[1-
D^{[111]}
\:\left(\frac{a_{\parallel}}{a^{epi}}-1
\right)\right]\\                   
\label{d111}
D^{[111]} &=& 2\: \left ( \frac{c_{11}+2\:c_{12}-2\:c_{44}}{c_{11}+2\:
c_{12}+4\:c_{44}} \right)\\ \nonumber
\epsilon^{epi}_{\parallel} & = & \frac{a_{\parallel}}{a^{epi}}-1 \\ \nonumber
\epsilon^{epi}_{\perp} & = & \frac{a_{\perp}^{epi}}{a^{epi}}-1
\end{eqnarray}
where $c_{ij}$ are the elastic constants for the bulk epilayer
(we have used the experimental values reported in Ref.\cite{harrison}).

Through total energy minimization of the ``ideal''
(AC)$_1$/(BC)$_1$ unrelaxed structures (in which all
the atoms are arranged in a zincblende structure with lattice constant $a$
-~our free parameter~-
and with bond-lengths $d_{AC}=d_{BC}=a\:(\sqrt{3}/4)$ ), we have found an
in-plane lattice  constant very close to the average of the bulk constituents,
according to Vegard's rule.
We have thus examined a free standing mode structure (indicated in the
following
as Elastically Relaxed or simply ER), that has this value for the
in-plane lattice constant. In Table \ref{parstruttan} and
Table \ref{parstruttcat}
we report the calculated structural parameters
for the ternary 
common-anion and  common-cation  systems:
the S1 (S2) system is a  common-anion SL grown on a
GaSb (InSb) substrate, while the S3 (S4) system is a common-cation
SL grown on an InAs (InSb) substrate.

In the case of pseudomorphic growth on a substrate,
we have found general agreement 
between the structures obtained through total energy minimization
and those given by MTE; this fully
justifies our having considered this approximation to
determine the five unknown parameters in the free standing mode (FSM) structure.
Note that deviations from the results predicted by MTE occur in the
case of InSb strained to GaSb or to InAs; in both these structures, total
energy
minimization gives an InSb bond-length that is
systematically larger (within 0.6 $\%$) than the one
expected according to MTE, even though the difference between the total energies
for the elastic SL and for the total energy minimal structures is very small
(barely larger than  the numerical uncertainty of 1 mRy/unit cell).
However, this can be justified considering that, due to
its elastic properties, this material could easily be out of
the linear elastic region.
In fact, the elastic constants for InSb are quite smaller than those for GaSb
and InAs \cite{harrison}, resulting in a larger effective strain -~due to
the mismatch~- in the former case; this is also confirmed by the non-linear
behaviour of the band-gap
as a function of the strain, as will be discussed later.
 
As expected from
the similarity of the GaSb and InAs elastic constants
\cite{harrison} and bulk moduli (the experimental values
are $B^{InAs}$ = 0.579 Mb and $B^{GaSb}$ = 0.578 Mb \cite{landbor}), we
obtain similar deformations for these two constituents
respectively in the common-anion and
common-cation SLs. We also notice that strains
(either parallel or perpendicular) and
percentage deviations from bulk bond-lengths are more pronounced in the
common-cation systems, compared to the common-anion systems: this is obviously
a consequence of the greater mismatch between the constituent lattice
parameters in the (InAs)$_1$/(InSb)$_1$ structures.
 
In the case of [001] ordered systems
(whose structural parameters are reported in Table \ref{parstruttan}
and Table \ref{parstruttcat} respectively for common-anion and common-cation
SL),
we obtain  a tetragonal
Bravais lattice with $D^5_{2d}$ 
space group \cite{cryst} and a
unit cell in real space with 4 atoms (two of which are equivalent);
the origin is taken on a cation (anion) site for the common-anion
(common-cation) system.
 
The  MTE relations reported in Eqs.(\ref{d111})
are still valid for [001] ordered SL,
if the parameter $D$ is redefined as:
\begin{equation}
D^{[001]} = 2\: \left ( \frac{c_{12}}{c_{11}}
\right). \nonumber
\end{equation}
 
A comparison between 
the [111] and [001] ordered structures 
(having the same chemical composition)
shows that in the same growth conditions (FSM or growth on a substrate)
the structural parameters are not equal.
In particular we notice larger perpendicular strains in the [001]
compared to the [111] ordered structures, while
 the parallel strains are obviously
equal in considering the same growth conditions; what we find is  thus
a smaller deviation from bulk bond-lengths,
due to a more effective relaxation.

\section{Electronic properties}
 
\subsection{Electronic levels}
 
The determination of the SL electronic energy levels is a fundamental
point for most of the properties of interest in the systems considered.
In Table \ref{sllivris} we report the calculated electronic
levels (with a numerical
uncertainty of $\pm$~0.04~eV, equal for all  the
energies reported in the present work, unless otherwise specified)
at the BZ center ($\Gamma$),
for the different [111] ordered systems considered
(free standing mode and pseudomorphic growth on the two substrates), both
for the common-anion and for the common-cation systems.
We also report the zincblende state from which
the SL state derives through folding the f.c.c. Brillouin zone back
into the smaller ternary trigonal zone.

The splitting, $\Delta_{CF}$,
of the triply degenerate $\Gamma_{15v}$ zincblende state
is due to the non-cubic crystal field and is conventionally taken
positive if the doubly degenerate state $\Gamma_{3v}^{(2)}$
has a higher energy compared to the state $\Gamma_{1v}$.
As can be seen from Table \ref{sllivris},
we obtain a negative $\Delta_{CF}$
in the case of GaSb  (InAs) strained to  InSb
for common-anion (cation) systems -
corresponding  to an in-plane extensive strain $\epsilon_{\parallel}$ -
whereas
the complementary case (pseudomorphic growth on a GaSb-substrate
(InAs-substrate))
and the free standing mode produce a positive $\Delta_{CF}$.
 
The introduction of spin-orbit coupling removes the double degeneracy of
the $\Gamma_{3v}^{(2)}$ state
and yields the electronic energy levels
illustrated in Fig.\ref{elenlev} as a function of
the substrate lattice constant. The topmost valence bands (E$_1$, E$_2$,
E$_3$) have been labeled according to the ``quasi-cubic" model \cite{hopf}
(taking the centre of gravity of the SL valence bands as zero);
 
\begin{center}
\begin{tabular}{lll}
E$_{1}$ & = & $+\frac{1}{3}(\Delta_{s.o.}+\Delta_{CF})$\\
E$_{2,3}$ & = & $-\frac{1}{6}(\Delta_{s.o.}+\Delta_{CF})\pm
\frac{1}{2}\{(\Delta_{s.o.}+\Delta_{CF})^2-\frac{8}{3}\:\Delta_{s.o.}\:
\Delta_{CF}\}^{1/2}$\\
\end{tabular}
\label{LEVELS}
\end{center}
considering $\Delta_{CF}$ as obtained from Table \ref{sllivris}
and $\Delta_{s.o.}$ for the SL as the value averaged over the equivalent
calculated \cite{ale}
quantities for the binary constituents 
(Even neglecting the $\Delta_{s.o.}$ negative bowing,
occurring in the common-cation SL \cite{az89}, it is possible to uniquely
identify the SL levels with the ``quasi-cubic" ones, which differ at most \
by 0.05~eV). From Fig.\ref{elenlev} we notice that the
E$_1$ and E$_c$ (the lowest conduction state)
levels show an almost linear behaviour
as a function of the 
substrate lattice constant, while the other valence band states
show a more complex trend, due to the interplay between crystal-field
and s.o. effects (see the cross-over between the E$_1$ and E$_2$ states).

In Table \ref{sllivris001} we report the relevant
electronic energy levels for the CuAu
systems at $\Gamma$; the notation is analogous to
that of Table \ref{sllivris}, where
folding relations of the f.c.c. Brillouin zone
in the  now  tetragonal ternary zone involve the
  zincblende states  as indicated in the Table.
The trend in the signs of the crystal-field splittings
$\Delta_{CF}$ is similar to that evidenced in the
[111] ordered systems (see Table \ref{sllivris001}).

The general underestimate of the band-gap energy in LDA
has stimulated many attempts to solve this  problem, but
 correction algorithms \cite{GW,hyb,SIC} need an
extraordinary computational effort in the SL case;
thus our LDA band-gap energy ($E_{gap}^{LDA}$)
was corrected starting
from the experimental data of the binary constituents \cite{az91}.
Due to a lack
of experimental band-gaps regarding strained binaries,
we have fitted the calculated values
obtained for each  binary 
in different strain
conditions ({\em i.e.} tetragonal and trigonal),
assuming a linear trend
for the band-gap energy as a function of
the in-plane strain $\epsilon_{\parallel}$:
$E_{gap}^{LDA}(\epsilon_{\parallel})
= E_{gap}^{LDA}(0) + \alpha\:\epsilon_{\parallel}$,
where $E_{gap}^{LDA}(0)$ is the
binary equilibrium  calculated band-gap
(a parabolic trend
$E_{gap}^{LDA}(\epsilon_{\parallel})
= E_{gap}^{LDA}(0) + \alpha\:\epsilon_{\parallel}
+ \beta\:\epsilon_{\parallel}^2$
has been used for InSb, which is assumed to be out
of the linear region).
Once we determined the coefficient $\alpha$, 
we translated the curve so that it becomes:
$E_{gap}^{emp}(\epsilon_{\parallel})
= E_{gap}^{expt}(0) + \alpha\:\epsilon_{\parallel}$,
where 
 $E_{gap}^{expt}(0)$ is the
binary equilibrium  experimental band-gap.
We have thus used these empirical values to
obtain the empirical  band-gap  energy averaged
over the strained binaries ($<E_{gap}^{emp}>$).
Summing this quantity to the correction
($\delta\:=\:E_{gap}^{LDA}(SL) - <E_{gap}^{LDA}>$ ), we have finally obtained
the predicted band-gap energy in the SL  ( $E_{gap}^{emp}(SL)$ )
with a numerical uncertainty of  $\pm$ 0.05~eV.
Although this procedure is empirical, it is expected to give good estimates of
the real band-gaps, since it is well known that while the band-gap is strongly
underestimated, the band-gap behaviour as a function of pressure is always
very well reproduced by LDA \cite{ale}.
 
We report in Tables \ref{slgap} and \ref{slgap001}
 (respectively for [111] and
[001] ordered structures),
the band-gap energies as obtained from LDA
self-consistent unperturbed
calculations ($E_{gap}^{unp}$), with the introduction of the
perturbation due to spin-orbit  coupling ($E_{gap}^{LDA}$)
and with the correction
starting from experimental data ($E_{gap}^{emp}$). From these Tables
we first notice that
for all  the systems considered
we find a negative
$E_{gap}^{LDA}$
(due to an inversion which causes the conduction band minimum (CBM) to lie
below the valence band maximum (VBM)). Furthermore, we observe that
the larger the substrate lattice parameter, the smaller
the band-gap energy (either in common-anion or in
common-cation systems).

A comparison between Table \ref{slgap} and Table \ref{slgap001}
confirms the trend predicted  by Wei and Zunger \cite{az91} for the energy
band-gap:
\begin{equation}
E_{gap}^{[111]}\:<\:E_{gap}^{[001]}\:<\:E_{gap}^{ave}
\label{andamento}
\end{equation}
where $E_{gap}^{[111]}$ and $E_{gap}^{[001]}$ are the band-gap energies
respectively  in the [111] and in the [001] ordered
structures, while
$E_{gap}^{ave}$ is the band-gap average energy taken over the 
binary constituents
(the calculated values of the LDA band-gap energy for the binary constituents
are $E_{gap}^{LDA}(GaSb)$ = - 0.47~eV,  $E_{gap}^{LDA}(InSb)$ = - 0.67~eV
and  $E_{gap}^{LDA}(InAs)$ = -~0.63 eV).  
 
As is well-known \cite{az89}, band folding in the superstructures
causes a repulsion between two binary electronic states
of different symmetries, folded on a state of the same symmetry in the ternary phase and
coupled through the  perturbative
potential mentioned in Section I (in Tables \ref{sllivris} and \ref{sllivris001}
the superscripts (1) and (2)
indicate the two states involved in the repulsion mechanism).
One of its interesting effects
is the band-gap narrowing,
compared to the equivalent quantity averaged over the binary
constituents (as confirmed by the second inequality in Eq.(\ref{andamento})).
The amount of this effect \cite{az89} is inversely proportional to the
difference
$[\epsilon(\Gamma_{1c})-\epsilon(L_{1c})]$ (in the [111]  structure)
or  to the difference $[\epsilon(\Gamma_{1c})-\epsilon(X_{1c})]$
(in the [001]  structure);   this difference is smaller
in the [111]  structure, causing
a more striking band-gap narrowing than in the [001] structure
(as shown by the first inequality in Eq.(\ref{andamento})).
These observations are confirmed by the calculated values for the
band-gap bowing
parameters (defined,  in analogy with the  50$\%$-50$\%$ alloys, as
$b_{gap}=4(E_{gap}^{ave}-E_{gap}$)) reported in Tables \ref{slgap}
and \ref{slgap001}:
we obtain a larger bowing in the [111] structures compared to the [001] ones
and, looking at  the  constituent chemical  species,
we can say that the bowing in
common-cation systems is larger than in the common-anion ones.
 
The band-gap trend as a function of the substrate lattice constant and its
dependence on the ordering direction have been illustrated in Fig.\ref{gap},
where we report the LDA band-gap ($E^{LDA}$~-~solid line)
and corrected  band-gap
($E^{emp}$~-~dashed line) as a function of
the substrate lattice constant for (GaSb)$_1$/(InSb)$_1$ (Fig.\ref{gap} (a))
and  (InAs)$_1$/(InSb)$_1$ (Fig.\ref{gap} (b)).
A comparison between the SL energy band-gaps and
the average value ($E_{ave}$)
of the experimental (LDA) band-gap in the pure binaries
-~indicated by the filled (empty) circles~- clearly shows the band-gap
narrowing
effect.
 
Tables \ref{slgap} and \ref{slgap001} also show that the
crystal field splittings in the [111] ordered structures
are always bigger than in the [001] structures in the  same growth conditions,
with the only exception  represented by
the S4 system, which has a smaller $\Delta_{CF}$ compared to the other structures.
This apparently strange behaviour
can be explained by considering that the
$\Gamma_{3v}^{(2)}$ state interacts with the lower
$\Gamma_{3v}^{(1)}$ state, resulting in an upward shift
(an effect relevant only in the common-cation SLs);
furthermore, in the S4 system, the VBM is a
$\Gamma_{1v}$ state, which is
only slightly involved in the level repulsion mechanism. Thus
the stronger the level repulsion, the larger the
$\Gamma_{3v}^{(2)}$ upward shift and the smaller
 $\Delta_{CF}$ becomes: this observation is thus a  further proof of
the validity of  the band repulsion model.
 
We observe that in the [111] ordered SLs the
band-gap energy is determined by the difference in energy between
the VBM -\uni\ slightly localized on the
anion belonging to the
InSb monolayer (as will be clearly shown in the next section)\uni\ -
and the CBM -\uni\ strongly localized
on the GaSb (InAs) sublattice in  the common-anion (cation) superlattice.
Thus we could think of  (GaSb)$_1$/(GaSb)$_1$-type systems as
common-anion SLs in which we substitute
the InSb monolayer  with a  GaSb monolayer.
We would expect in this case a small
modification  of the band-gap energy, since the VBM will no longer be
localized on the Sb belonging to the InSb sublattice but rather
on the Sb
belonging to a GaSb sublattice; therefore, this will be only 
a second order effect.
Our prediction is confirmed by the
calculated band-gap energies
(spin-orbit included) reported in
Table \ref{gapcoman} (second column)
which prove the almost total independence of the VBM on the cationic
substitution;
the change of the band-gap energy in the different structures is thus caused
by the structural
term in the perturbative SL potential rather than by the chemical term.
An equivalent interpretation considers
the (InSb)$_1$/(InSb)$_1$-type systems as SLs in which
we have substituted the GaSb monolayer
(where the CBM is strongly localized)
with an InSb monolayer.
In this case, the cationic substitution  implies the chemical
alteration of one of the atomic species (Ga) on which the wave function is
strongly localized; thus, what we expect, is an appreciable change in the
band-gap energy, as confirmed by the
third column in Table \ref{gapcomcat} (from which we notice the
 $E_{gap}^{LDA}$ increase).

An analogous trend is observed for the common-cation systems, where InAs
has now susbstituted the GaSb
as the InSb partner in the SL (see second and third column in Table
\ref{gapcomcat}).
 
The trend in the crystal field splitting is strongly dependent on the
class of systems
considered. In fact, in the common-anion systems, this quantity is almost
independent of the cationic substitution (as  expected, because of the
 anionic character of the VBM, localized
on the Sb atom).
In this case, the chemical term of the potential existing in the SL
has very little effect on $\Delta_{CF}$, compared to the structural term
(see Table \ref{gapcoman}). On the other hand,  in common-cation systems
anionic substitution has a strong effect on the crystal field
splitting  and the chemical term in the SL
potential is now much more important than before,
even though
the structural term still has a strong effect on $\Delta_{CF}$
(see Table \ref{gapcomcat}).

\subsection{Charge density distribution}
 
One of the main effects of the perturbative potential  in the SL
(in particular of its chemical term, due to differences in the constituent
atom's orbital energies \cite{az94}) is the
localization of the charge density
in one of the constituent monolayers, which  varies from state to state.
As an obvious consequence, this effect causes the confinement of the charge
carriers (holes or electrons  respectively for valence or conduction
states) in a different sublattice.
 
In order to better clarify the character relative to
the different states of interest, their angular decomposition 
-~for the common-anion
systems in the three  growth conditions considered~-
is reported in Table \ref{decoman}
(we do not to report the equivalent Table for
(InAs)$_1$/(InSb)$_1$,
since this system is very similar to the previous one, as far as the charge
decomposition is concerned).
Referring to the charge density of the $\Gamma_{1v}$ state,
we notice in particular the growing $s$ character
on the InSb monolayer and the decreasing $p$ character on the In atom
as the substrate lattice parameter is increased;
 at the same time, the $s$ charge density on the GaSb sublattice decreases,
 while the $p$ charge grows on the Ga atom.
 
We report in Fig.\ref{gammav} (a) distribution for the $\Gamma_{1v}$ state for
the (GaSb)$_1$/(InSb)$_1$ 
elastically relaxed (ER) systems,
drawn the same for all the
charge densities reported in this work in a plane
perpendicular to the atomic layers.
This state comes from $p_z$ orbitals (as we can see from the typical
``butterfly'' shape along the vertical growth $z$-direction)
and shows a strong bonding character, between different
monolayers and within each monolayer.

In Fig.\ref{gammav} (b) 
we  report  the charge density distribution
relative to the  $\Gamma_{3v}^{(2)}$
(VBM) state for the common-anion system
in its elastically relaxed structure, where the  localization
of the charge density in the InSb sublattice
is particularly evident. We notice that this peculiarity
is much more enhanced in the common-cation
system (not shown), as a probable
consequence of the anionic character of
this state: in fact, what we expect  in the common-cation system is
for the Sb-atom to draw more charge than the As-atom.
We have found that the charge density distribution in this state
is not strongly influenced by the strain conditions, as can be seen from
Table \ref{decoman}.
 
The calculated charge density distribution
for the first conduction state  $\Gamma_{1c}^{(1)}$
(CBM), relative to the common-anion elastically relaxed system,
is presented in Fig.\ref{gammac} (a).
What is relevant in this state is the strong localization
of the charge density in the
GaSb monolayer (a similar
behaviour is shown by the common-cation ER system, where the charge
density is concentrated on the InAs monolayer).
 
The localization emphasized above becomes more pronounced as
the substrate lattice  parameter is increased:
the charge density distribution concentrates more and more
in the GaSb monolayer (InAs monolayer) while
at the same time the InSb monolayer
becomes charge-depleted (as confirmed by Table \ref{decoman}).
The second conduction state, $\Gamma_{1c}^{(2)}$,
shows a complementary trend,
owing to the charge density distribution that is more and more
concentrated on the InSb sublattice as the substrate lattice
parameter is increased
(as we notice from Fig.\ref{gammac} (b) for
the common-anion  elastically relaxed system; this behaviour is
similar
to the common-cation systems).
 
As a consequence of these observations, in all these structures we have
a direct gap in reciprocal space, while we obtain a
``spatially indirect'' gap, due to the localization of the
$\Gamma_{3v}^{(2)}$ state (VBM) on the InSb sublattice
and of the $\Gamma_{1c}^{(1)}$ state (CBM) on the GaSb
(InAs) sublattice in the common-anion (common-cation) SL.
 
\section{Conclusions}
 
{\em Ab initio} FLAPW calculations, based on density functional theory
within LDA, have been performed in order to determine the
electronic properties of ultrathin SLs. In particular we have studied
a common-anion  (GaSb)$_1$/(InSb)$_1$ system and a common-cation
 (InAs)$_1$/(InSb)$_1$ system, ordered along  two different ([111] and [001])
directions .
 
The relevant results obtained for these structures can be summarized
as follows:
 
\begin{enumerate}
 \item Both the (GaSb)$_1$/(InSb)$_1$ and  (InAs)$_1$/(InSb)$_1$ systems
show a direct gap ($E_{gap}^{[111]}$) which is smaller than the average
band-gap energy ($E_{gap}^{ave}$) taken over the binary constituents: the
dependence of this quantity on the ordering direction is expressed by the
relation:
$E_{gap}^{[111]}\:<\:E_{gap}^{[001]}\:<\:E_{gap}^{ave}$;
 
\item Both common-anion and common-cation systems show a decreasing
band-gap energy as the substrate lattice parameter is increased;
 
\item The structures studied offer interesting opportunities for band-gap
tuning as a function of growth conditions;
the range in which the gap varies is as large as 0.7~eV
in    (GaSb)$_1$/(InSb)$_1$ type systems and  0.3~eV
 in   (InAs)$_1$/(InSb)$_1$ systems;
 
\item In the common-anion (common-cation) systems
the marked charge density localization of the CBM on the GaSb (InAs)
monolayer and of the VBM on the InSb monolayer causes the gap to be
``spatially indirect'';
 
\item In the case of
free standing mode elastically relaxed structures we obtain a band-gap
value of 0.05~$\pm$~0.05~eV in the common-anion system
(semiconducting properties) and of -0.26~$\pm$~0.05~eV
in the common-cation system (semimetallic properties);
 
\end{enumerate}
 
\section{ACKNOWLEDGEMENTS}
We thank B.W. Wessels for stimulating discussions and a
careful reading of the manuscript.
Work at Northwestern University supported by the MRL Program of the
National Science Foundation, at the Materials Research Center of
Northwestern University, under Award No. DMR-9120521, and by a grant of
computer time at Pittsburgh Supercomputing Center.

\newpage

 
\begin{table}
\centering
\caption{Bond-lengths ($d_{GaSb}$ and $d_{InSb}$ in a.u.)
and strain parameters ($\epsilon^{GaSb}$
and $\epsilon^{InSb}$) parallel and perpendicular to the growth plane
 for  (GaSb)$_1$/(InSb)$_1$
[111] and [001] ordered systems. The quantities denoted by $\Delta$
indicate  percentage deviations from calculated bulk bond-lengths
($d_{GaSb}$ = 5.00 a.u. and $d_{InSb}$ = 5.29 a.u.).}
\vspace{5mm}
\begin{tabular}{|c|ccc|ccc|}
& & [111] & & & [001] & \\
& El.Rel. & GaSb-subs. & InSb-subs.&
 El.Rel. & GaSb-subs. & InSb-subs.\\ \hline \hline
$d_{GaSb}$ & 5.07  &  5.00  & 5.15 & 5.05 & 5.00 & 5.11\\
$\Delta_{GaSb}$ &+1.4 \% & - & +3.0 \% & +1.0 \%& - & 2.2 \%\\
$\epsilon_{\parallel}^{GaSb}$ & +0.029 & 0 & +0.058 & +0.029 & 0 & +0.058
\\      
$\epsilon_{\perp}^{GaSb}$ & -0.014 & 0 & -0.028 & -0.027 & 0 & -0.053 \\
\hline \hline
 $d_{InSb}$ & 5.22  &  5.16  & 5.29 & 5.25  & 5.24  & 5.29  \\
$\Delta_{InSb}$&  -1.3 \% & -2.5 \% & - & -0.8 \% & -0.9 \% & -\\
$\epsilon_{\parallel}^{InSb}$ & -0.028 & -0.055 & 0 & -0.028 & -0.055 & 0
\\      
$\epsilon_{\perp}^{InSb}$ & +0.016 & +0.033 & 0 & +0.030 & +0.076 & 0 \\
\end{tabular}
\label{parstruttan}
\end{table}
 
\begin{table}
\centering
\caption{Bond-lengths ($d_{InAs}$ and $d_{InSb}$ in a.u.)
and strain parameters ($\epsilon^{InAs}$
and $\epsilon^{InSb}$)  parallel and perpendicular to the growth plane
 for  (InAs)$_1$/(InSb)$_1$
[111] and [001] ordered systems. The quantities denoted by $\Delta$
indicate  percentage deviations from calculated bulk bond-lengths
($d_{InAs}$ = 4.96 a.u. and $d_{InSb}$ = 5.29 a.u.).}
\vspace{5mm}
\begin{tabular}{|c|ccc|ccc|} 
& & [111] & & & [001] & \\
& El.Rel. & InAs-subs. & InSb-subs. & El.Rel. & InAs-subs. & InSb-subs. \\
\hline \hline
$d_{InAs}$ & 5.04  &  4.96  & 5.12 & 5.01 & 4.96 & 5.07\\
$\Delta_{InAs}$ &+1.6 \% & - & +3.2 \% & +1.0 \%& - & 2.2 \%\\
$\epsilon_{\parallel}^{InAs}$ & +0.033 & 0 & +0.066 & +0.033 & 0 & +0.066
\\      
$\epsilon_{\perp}^{InAs}$ & -0.019 & 0 & -0.038 & -0.036 & 0 & -0.072 \\      \hline \hline
 $d_{InSb}$ & 5.21  &  5.14  & 5.29 & 5.24  & 5.22  & 5.29  \\
$\Delta_{InSb}$&  -1.5 \% & -2.8 \% & - & -0.9 \% & -1.3 \% & -\\
$\epsilon_{\parallel}^{InSb}$ & -0.031 & -0.062 & 0 & -0.031 & -0.062 & 0
\\      
$\epsilon_{\perp}^{InSb}$ & +0.019 & +0.037 & 0 & +0.033 & +0.078 & 0 \\
\end{tabular}
\label{parstruttcat}
\end{table}

\begin{table}
\centering
\caption{Calculated electronic energy levels (in eV) with respect to the VBM
for 
the [111] SL (neglecting s.o. coupling).
The superscripts (1) and (2)
indicate the two states involved in the repulsion mechanism.
The state multiplicity is given in parentheses.}
\vspace{5mm}
\begin{tabular}{|c|c|ccc|ccc|} 
SL State & ZB State & 
&  (GaSb)$_1$/(InSb)$_1$ &  &
& (InAs)$_1$/(InSb)$_1$ &  \\
&& ER & S1 & S2 & ER & S3 & S4\\\hline \hline
 $\Gamma_{1c}^{(2)}$ & $L_{1c}$ & 0.50 & 0.70 & -0.20
& 0.79 & 1.04 & 0.36 \\
 $\Gamma_{1c}^{(1)}$ & $\Gamma_{1c}$ & -0.81 & -0.64 & -1.47
 & -1.02& -0.95 & -1.42\\
         & & &         &  &       &       &     \\
 $\Gamma_{3v}^{(2)}$ &$\Gamma_{15v}$ & 0(2) & 0(2) &  -0.42(2)
 & 0(2) & 0(2)& -0.18(2)  \\
$\Gamma_{1v}$ & $\Gamma_{15v}$& -0.11  & -0.59  &  0
& -0.21  & -0.77& 0    \\
 $\Gamma_{3v}^{(1)}$ & $L_{3v}$ & -1.31(2) & -1.35(2) & -1.68(2)
 & -1.38(2)  & -1.46(2)&
 -1.50(2) \\ 
\end{tabular}
\label{sllivris}
\end{table}
 
\begin{table}
\centering
\caption{Calculated electronic energy levels (in eV) with respect to the VBM
for 
the [001] SL (neglecting s.o. coupling).
The superscripts (1) and (2)
indicate the two states involved in the repulsion mechanism.
The state multiplicity is given in parentheses.}
\vspace{5mm}
 
\begin{tabular}{|c|c|ccc|ccc|} 
SL State & ZB State & 
& (GaSb)$_1$/(InSb)$_1$ &  &
& (InAs)$_1$/(InSb)$_1$ & \\
&& ER & S1 & S2 & ER & S3 & S4\\\hline \hline
 $\Gamma_{1c}^{(2)}$ & $X_{1c}$ & 1.12& 1.26 & 0.78
& 1.46 & 1.54 & 1.01 \\
 $\Gamma_{1c}^{(1)}$ & $\Gamma_{1c}$ & -0.49 & -0.43 & -0.77
 & -0.68 & -0.69 & -0.96\\
         & & &         &  &       &       &     \\
 $\Gamma_{5v}^{(2)}$ &$\Gamma_{15v}$ & 0(2) & 0(2) &  -0.20(2)
 & 0(2) & 0(2)& -0.25(2)  \\
$\Gamma_{4v}$ & $\Gamma_{15v}$& -0.07  & -0.41  &  0
& -0.06  & -0.46 & 0   \\
 $\Gamma_{5v}^{(1)}$ & $X_{5v}$ & -2.52(2) & -2.52(2) & -2.71(2)
 & -2.46(2)  & -2.46(2)& -2.70(2) \\ 
\end{tabular}
\label{sllivris001}
\end{table}

\begin{table}
\centering
\caption{Band-gap  energies (in eV) for
(GaSb)$_1$/(InSb)$_1$ [111] systems and (InAs)$_1$/(InSb)$_1$ [111]
obtained from unperturbed LDA
calculations ($E_{gap}^{unp}$), with the introduction of spin-orbit coupling
($E_{gap}^{LDA}$) and corrected starting from experimental data
($E_{gap}^{emp}$). We also report the calculated
bowing parameter ($b_{gap}^{[111]}$) for the different
systems considered.}
\vspace{5mm}
\begin{tabular}{|c|ccc|ccc|} 
\centering
& &  (GaSb)$_1$/(InSb)$_1$ & &
& (InAs)$_1$/(InSb)$_1$ & \\
 & ER & S1 & S2 & ER & S3 & S4 \\ \hline \hline
$E_{gap}^{unp}$  &-0.81& -0.64 & -1.47 & -1.02 & -0.95 & -1.42\\
$E_{gap}^{LDA}$ & -1.05& -0.90 & -1.59 & -1.26 & -1.20 & -1.49\\
$E_{gap}^{emp}$ & 0.05& 0.20 & -0.47 & -0.26 & -0.21 & -0.51\\ \hline
$b_{gap}^{[111]}$ & 1.92 & 1.32 & 4.08 & 2.42 & 2.18 & 3.34 \\ 
\end{tabular}
\label{slgap}
\end{table}
 
\begin{table}
\centering
\caption{Band-gap  energies (in eV) for
(GaSb)$_1$/(InSb)$_1$ [001] systems and (InAs)$_1$/(InSb)$_1$ [001]
obtained from unperturbed LDA
calculation ($E_{gap}^{unp}$), with the introduction of spin-orbit coupling
($E_{gap}^{LDA}$) and corrected starting from experimental data
($E_{gap}^{emp}$). We also report the calculated
bowing parameter ($b_{gap}^{[001]}$) for the different
systems considered.}
\vspace{5mm}
\begin{tabular}{|c|ccc|ccc|} 
\centering
 & & (GaSb)$_1$/(InSb)$_1$ & & & (InAs)$_1$/(InSb)$_1$ & \\
 & ER & S1 & S2 & ER & S3 & S4 \\  \hline \hline
$E_{gap}^{unp}$  &-0.49 & -0.43 & -0.77 & -0.68 & -0.69 & -0.96\\
$E_{gap}^{LDA}$ & -0.74 & -0.69 & -0.96 & -0.88 & -0.89 & -1.07\\
$E_{gap}^{emp}$ & 0.33 & 0.41 & 0.13 & 0.11 & 0.10 & -0.09\\ \hline
$b_{gap}^{[001]}$ & 0.68 & 0.48 &1.56 &0.90 & 0.94 & 1.66 \\ 
\end{tabular}
\label{slgap001}
\end{table}

\begin{table}
\centering
\caption{Band-gap  energies  $E_{gap}^{LDA}$,  spin-orbit coupling
included,
and crystal-field splittings $\Delta_{CF}$ (in eV) for common-anion systems.}
\vspace{5mm}
\begin{tabular}{|c|ccc|ccc|ccc|} 
\centering
 & & (GaSb)$_1$/(InSb)$_1$ & & & (GaSb)$_1$/(GaSb)$_1$ & & &
(InSb)$_1$/(InSb)$_1$ & \\
 &ER & S1 & S2 & ER & S1 & S2 & ER & S1 & S2\\\hline  \hline 
$E_{gap}^{LDA}$  &-1.05& -0.90 & -1.59 & -1.15
& -1.14 & -1.66 & -0.34 & -0.26 & -0.88\\
$\Delta_{CF}$ & 0.11 & 0.59 & -0.42 & 0.19 & 0.58 & -0.42 & 0.14
& 0.57 & -0.50 \\ 
\end{tabular}
\label{gapcoman}
\end{table}
 
\begin{table}
\centering
\caption{Band-gap  energies  $E_{gap}^{LDA}$, spin-orbit coupling included,
and crystal-field splittings $\Delta_{CF}$ (in eV) for common-cation systems.}
\vspace{5mm}
\begin{tabular}{|c|ccc|ccc|ccc|} 
\centering
 & & (InAs)$_1$/(InSb)$_1$ & & & (InAs)$_1$/(InAs)$_1$ & & &
(InSb)$_1$/(InSb)$_1$ &  \\
 &ER & S3 & S4 & ER & S3 & S4 & ER & S3 & S4\\\hline \hline 
$E_{gap}^{LDA}$  &-1.26& -1.20 & -1.49 & -1.26 & -1.24 & -1.44 & -0.29 & -0.25 &
-1.09 \\
$\Delta_{CF}$ & 0.21 & 0.77 & -0.18 & 0.14 & 0.57 & -0.39 & 0.11
& 0.68 & -0.47 \\ 
\end{tabular}
\label{gapcomcat}
\end{table}

\begin{table}
\centering
\caption{Angular decomposition relative to the muffin tin charge density
(for $s$  ($Q_s$) and $p$ ($Q_p$) components)
 of the different states,
neglecting s.o. coupling, for the (GaSb)$_1$/(InSb)$_1$ [111] systems.}
\vspace{5mm}
\small
\begin{tabular}{|cc|cccc|cccc|cccc|} 
 & & & S1 & & & & ER & & & &  S2 & & \\
State& & $Ga$ & $Sb$ & $In$ & $Sb$ & $Ga$ & $Sb$ & $In$ & $Sb$ &
$Ga$ & $Sb$ & $In$ & $Sb$\\ \hline           \hline
 & & & & & & & & & & & & & \\
$\Gamma_{1v}^{}$ & $Q_s$ & 0.065 & 0.057 & 0.006 & 0.005 &
0.003 & 0.001 & 0.014 & 0.017 & 0.007 & 0.001 & 0.045 & 0.023\\
& $Q_p$ & 0.037 & 0.187 & 0.060 & 0.134 & 0.040 & 0.218 & 0.048 & 0.184 &
0.074 & 0.115 & 0.014 & 0.241\\
 & & & & & & & & & & & & & \\
$\Gamma_{3v}^{(2)}$ & $Q_s$ & 0.000 & 0.000 & 0.000 & 0.000 &
0.000 & 0.000 & 0.000 & 0.000 & 0.000 & 0.000 & 0.000 & 0.000\\
& $Q_p$ & 0.023 & 0.122 & 0.061 & 0.321 & 0.025 & 0.117 & 0.063 & 0.306 &
0.026 & 0.114 & 0.063 & 0.294\\ 
 & & & & & & & & & & & & & \\
 
$\Gamma_{1c}^{(1)}$ & $Q_s$ & 0.262 & 0.177 & 0.074 & 0.072 &
0.345 & 0.232 & 0.047 & 0.045 & 0.349 & 0.217 & 0.031 & 0.030\\
& $Q_p$ & 0.019 & 0.027 & 0.002 & 0.063 & 0.005 & 0.019 & 0.006 & 0.001 &
0.011 & 0.031 & 0.005 & 0.001\\ 
 & & & & & & & & & & & & & \\
 
$\Gamma_{1c}^{(2)}$ & $Q_s$ & 0.028 & 0.008 & 0.188 & 0.161 &
0.008 & 0.000 & 0.228 & 0.180 & 0.000 & 0.003 & 0.223 & 0.180\\
& $Q_p$ & 0.073 & 0.024 & 0.000 & 0.025 & 0.057 & 0.006 & 0.001 & 0.052 &
0.001 & 0.106 & 0.033 & 0.004\\   
\end{tabular}
\vspace{5mm}
\normalsize
\label{decoman}
\end{table}

\begin{figure}
\centering
\caption{
Calculated highest valence band energy levels (E$_1$, E$_2$ and
E$_3$) and lowest conduction state (E$_c$) at $\Gamma$
versus substrate lattice constant
for the [111] (a) common-anion  and (b) common-cation  SLs considered.
The center of gravity of the topmost valence bands is taken as zero.}
\label{elenlev}
\end{figure}
 
\begin{figure}
\centering
\caption{
LDA band-gap ($E^{LDA}$~-~solid line)
and corrected band-gap ($E^{emp}$~-~dashed line)  as a function of
the substrate lattice constant for the [111] and [001]
(a) (GaSb)$_1$/(InSb)$_1$ SLs and  (b) (InAs)$_1$/(InSb)$_1$ SLs.
Filled (empty) circles indicate
the average value ($E_{ave}$)
of the experimental (LDA) band-gap in the pure binaries.}
\label{gap}
\end{figure}
 
\begin{figure}
\centering
\caption{Charge density distribution
(in units of 0.5 e/unit cell)
for the (a)  $\Gamma_{1v}$
state and for the   (b) $\Gamma_{3v}^{(2)}$ state
in the elastically relaxed [111] ordered  common-anion structure.}
\label{gammav}
\end{figure}
 
\begin{figure}
\centering
\caption{Charge density distribution
(in units of 0.5 e/unit cell)
for the  (a)
$\Gamma_{1c}^{(1)}$
(CBM) state and for the (b)
  $\Gamma_{1c}^{(2)}$ state in the elastically relaxed
 [111] ordered  common-anion structure.}
\label{gammac}
\end{figure}


\begin{thebibliography}{100}
 
\bibitem{dj} D.J.Arent {\em et al.}, Appl. Phys. Lett. {\bf 62}, 1806 (1993).
 
\bibitem{laks} S.H.Wei, B.Laks and A.Zunger, Appl. Phys. Lett. {\bf 62}, 1937
(1993).
 
\bibitem{lemos} V.Lemos, C.Vasquez-Lopez and F.Cardeira, Superlatt. and
Microst.  {\bf 13}, 1891 (1993).
 
\bibitem{aleoff} A.Continenza, S.Massidda and A.J.Freeman, Phys. Rev. B
{\bf 42}, 3469 (1990).
 
\bibitem{christensen} N.E.Christensen and I.Gorczyka, Phys. Rev. B {\bf 50},
4397 (1994).
 
\bibitem{anderson} N.G.Anderson and S.Jones, J. of Appl.Phys. {\bf 70}, 4342
(1991).
 
\bibitem{tit} N.Tit, M.Peressi and S.Baroni, Phys. Rev. B {\bf 48}, 17607
(1993).
 
\bibitem{rubio} A.Rubio, J.L.Corkill and M.L.Cohen, Phys. Rev. B {\bf 49},
1952 (1994).
 
\bibitem{az94} A. Franceschetti, S.H.Wei and A.Zunger,
Phys. Rev. B  {\bf 50}, 8094 (1994).
 
\bibitem{vlaev} S.Vlaev, R.V.Velasco and F.Garc\'ia Moliner, Phys. Rev. B
{\bf 50}, 4577 (1994).
 
\bibitem{prins} A.D.Prins {\em et al.}, Phys. Rev. B {\bf 47}, 2191 (1993).
 
\bibitem{qw3} L.Q.Qian and B.W.Wessels, J. of Appl. Phys. {\bf 75},
3024 (1993).
 
\bibitem{materials} ``{\em Materials for Infrared Detectors and Sources}",
R.Farrow, J.F.Schetzina and J.J.Cheung (Materials Research Society, Pittsburg.
1987), Vol.90.
 
\bibitem{qwapl} L.Q.Qian and B.W.Wessels, Appl. Phys. Lett. {\bf 63},
628 (1993).
 
\bibitem{qwjvc} L.Q.Qian and B.W.Wessels,
J. Vac. Sci. Technol. {\bf 11} (4)., 1652 (1993).
 
\bibitem{inassb1} S.Elies, A.Krier, I.R.Cleverley and K.Singer,
J. Phys. D: Appl. Phys. {\bf 26}, 159 (1993).
 
\bibitem{inassb2} S.R.Kurtz, L.R.Dawson, R.M.Biefeld, D.M. Follstaedt and
B.L.Doyle, Phys. Rev. B {\bf 46}, 1909 (1992).
 
\bibitem{ohler} C.Ohler {\em et al.}, Phys. Rev. B {\bf 50}, 7833 (1994).
 
\bibitem{bensaada} A.Bensaada
{\em et al.}, J. of Appl. Phys. {\bf 75}, 3024 (1993).
 
\bibitem{VDW} C.G.Van de Walle, Phys. Rev. B {\bf 39}, 1871 (1988).
 
\bibitem{osscupt} For a review see A.Zunger and
S.Mahajan in ``{\em Handbook of
semiconductors}'', 2nd. ed., edited by S.Mahajan (Elsevier, Amsterdam, 
in press),
Vol.3 (and references therein). 

\bibitem{az89} S.H.Wei and A.Zunger, Phys. Rev. B {\bf 39}, 3279 (1989).
 
\bibitem{HK64} P.Hohenberg and W.Kohn, Phys. Rev. {\bf 136}, B864 (1964).;
W.Kohn and L.J.Sham, Phys. Rev. {\bf 140}, A1133 (1965).
 
\bibitem{HL}  L.Hedin and B.I.Lundqvist, J. Phys. C. {\bf 4}, 2064 (1971).
 
\bibitem{FLAPW}
H.J.F.Jansen and A.J.Freeman, Phys. Rev. B {\bf 30}, 561 (1984).;
M.Weinert, H.Krakauer, E.Wimmer and A.J.Freeman,
Phys. Rev. B {\bf 24}, 864 (1981).
 
\bibitem{MP} H.J.Monkhorst and J.D.Pack, Phys. Rev. B  {\bf 13}, 5188 (1976).
 
\bibitem{broy} C.G.Broyden, Math.Comp. {\bf 19}, 577 (1965).
 
\bibitem{cryst} ``{\em The Mathematical Theory of Symmetry in Solids}", C.J.
Bradley and A.P.Cracknell (Clarendon, Oxfords, 1972); ``{\em International
Tables for Crystallography}", J.Hahn (Reidel, Dordrecht, 1983), Vol.A.
 
\bibitem{magri}  R.Magri and C.Calandra, Phys. Rev. B {\bf 40}, 3896 (1989).
 
\bibitem{harrison}  ``{\em Electronic Structure and
the Properties of Solids}'', W.A.Harrison
(Freeman, San Francisco, 1980).
 
\bibitem{landbor} Landolt-B\"ornstein: ``{\em Numerical Data and Functional
Relationships in Science and Technology}'', Vol.17a
(Springer-Verlag, Berlin, 1982).
 
\bibitem{hopf} J.J.Hopfield, J. Phys. Chem. Solids {\bf 15}, 97 (1960).
 
\bibitem{ale} S.Massidda {\em et al.}, 
Phys. Rev. B {\bf 41}, 12079 (1990).
 
\bibitem{GW}
 L.Hedin, Phys. Rev. {\bf 139}, A796 (1965).
 
\bibitem{hyb} M.S.Hybertsen and S.G.Louie, Phys. Rev. Lett. {\bf 55}, 1418
(1985).
 
\bibitem{SIC}
 A.Svane and O.Gunnarsson, Phys. Rev. Lett. {\bf 65}, 1148 (1990).
 
\bibitem{az91} S.H.Wei and A.Zunger, Appl. Phys. Lett. {\bf 58}, 2685 (1991).
 
 
\end{thebibliography}
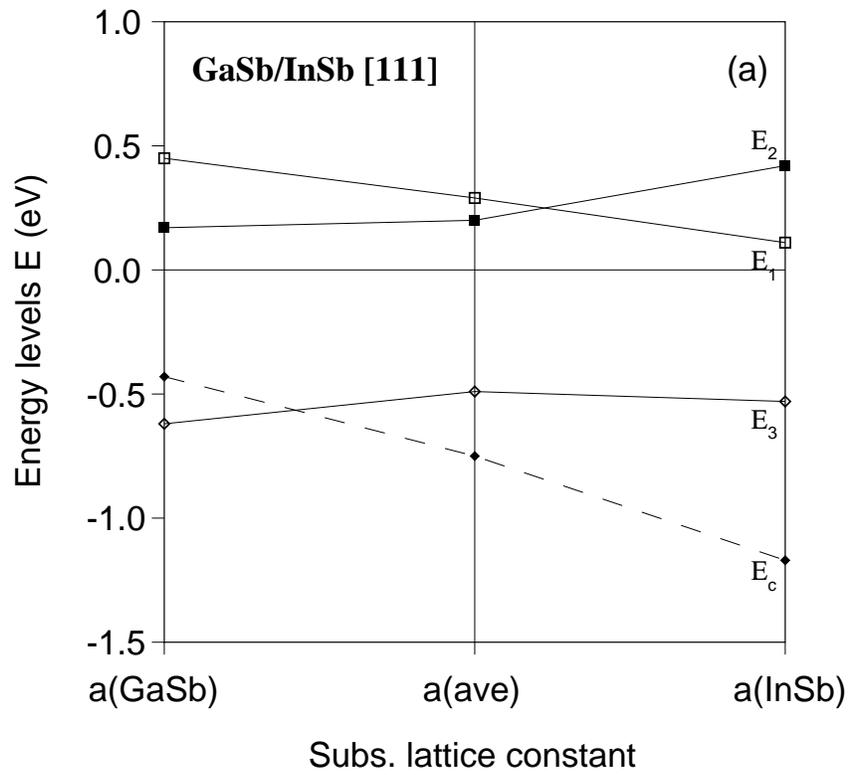
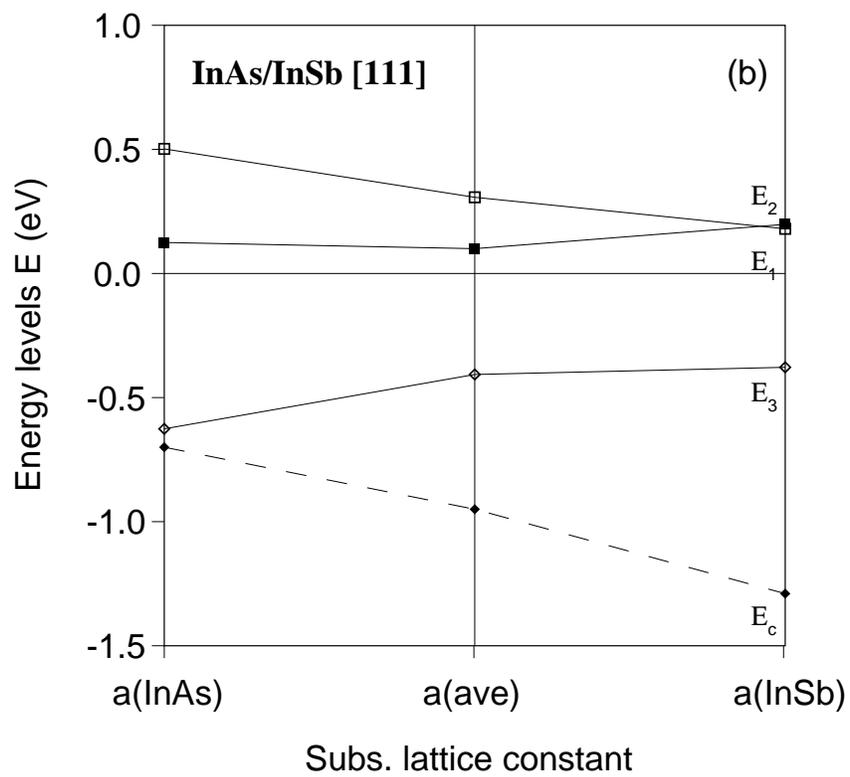
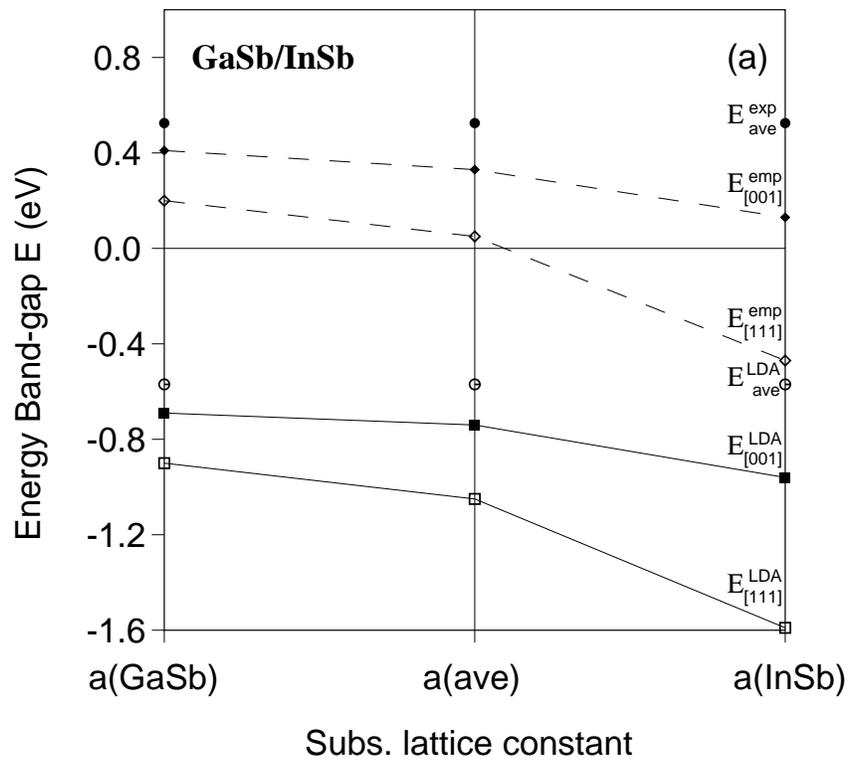
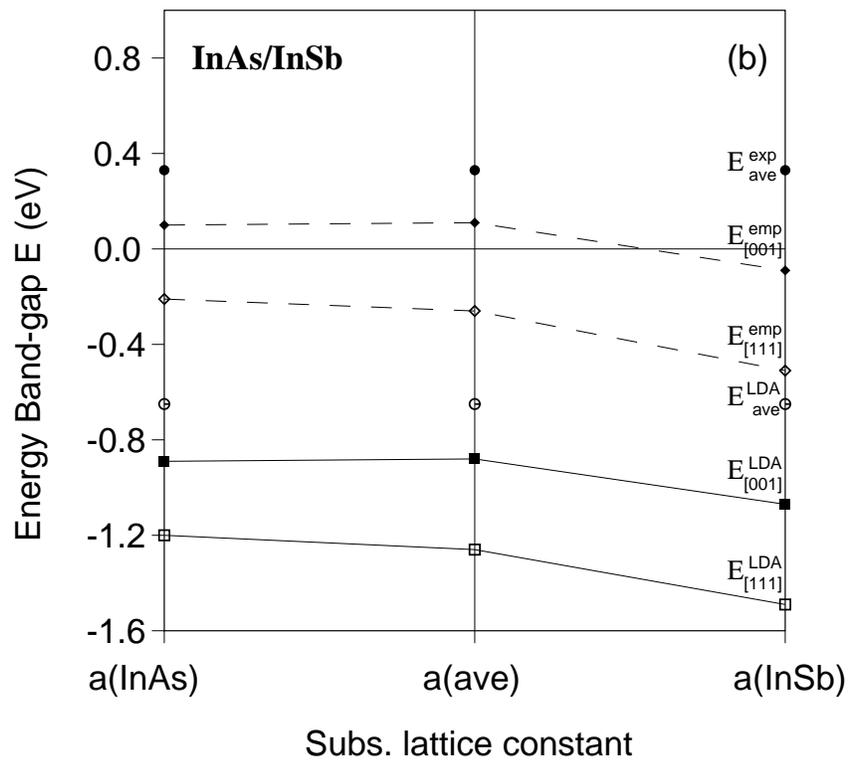
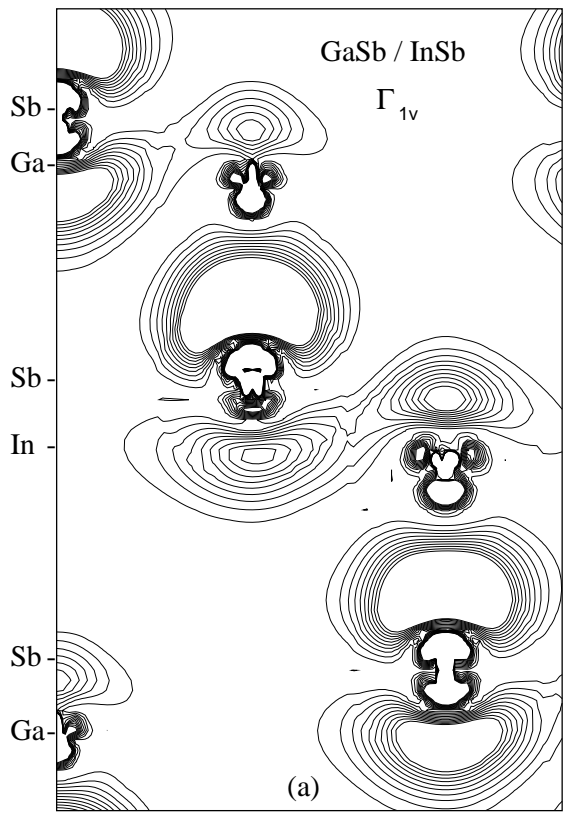
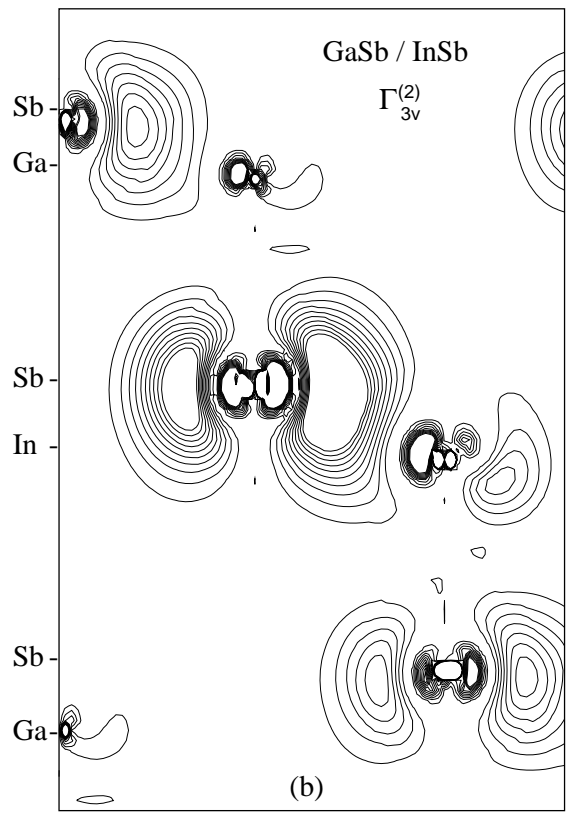
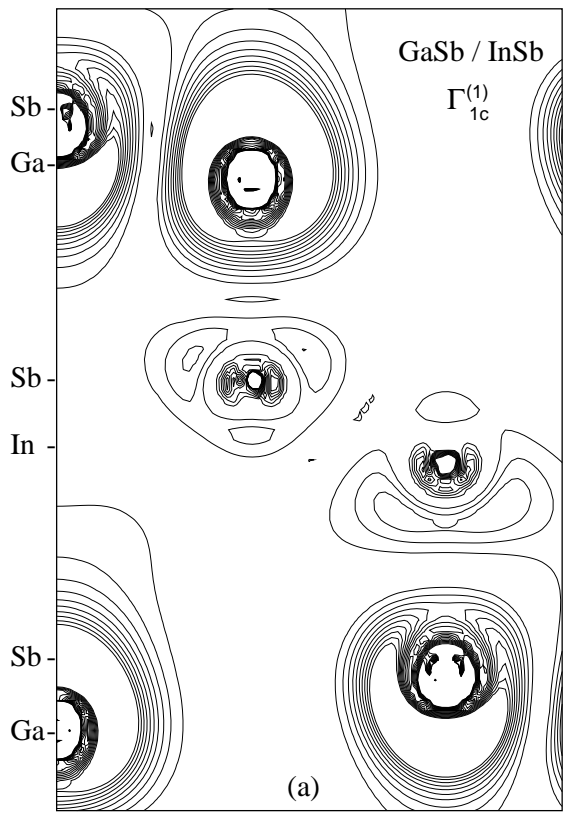
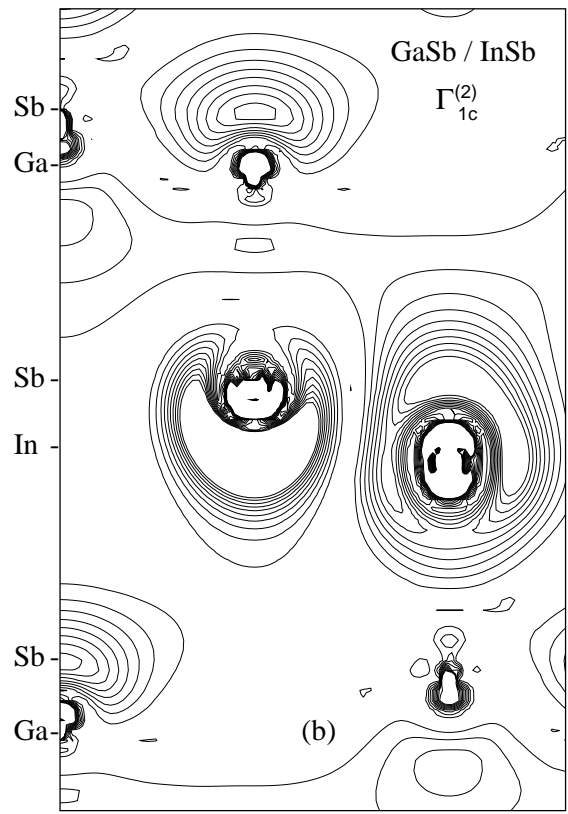
\end{document}